\documentclass[letterpaper, 10 pt, conference]{cssconf}

\IEEEoverridecommandlockouts                              

\usepackage{graphicx}

\def\theequation{\thesection.\arabic{equation}}

\newcommand{\be}{\begin{equation}}
\newcommand{\ee}{\end{equation}}
\newcommand{\bd}{\begin{displaymath}}
\newcommand{\ed}{\end{displaymath}}
\newcommand{\bea}{\begin{eqnarray}}
\newcommand{\eea}{\end{eqnarray}}

\newcommand{\Hi} {{\cal H}^\infty}

\newcommand{\compleib}{COMPL$_e$IB}

\def\matlab{{\sc \mbox{matlab}}}
\def\hifoo{{\sc \mbox{hifoo}}}
\def\hanso{{\sc \mbox{hanso}}}
\def\mosek{{\sc \mbox{mosek}}}
\def\slicot{{\sc \mbox{slicot}}}

\title{\LARGE \bf Fixed-Order $\Hi$ Controller Design via HIFOO,
a Specialized Nonsmooth Optimization Package}
\author{Suat Gumussoy\thanks{Courant Institute of Mathematical Sciences, New York University}
\and Michael L. Overton\thanks{Courant Institute of Mathematical Sciences, New York University}}
\date{}

\begin{document}
\maketitle

\begin{abstract}
We report on our experience with fixed-order $\Hi$ controller design
using the \hifoo\ toolbox.   We applied \hifoo\ to various benchmark fixed (or reduced)
order $\Hi$ controller design problems in the literature, comparing the results
with those published for other methods.
The results show that \hifoo\ can be used as an effective alternative to
existing methods for fixed-order $\Hi$ controller design.\end{abstract}

\section{Introduction}
In this note, we report on our experience applying \hifoo\
\cite{BHLO-IFAC-06} ($\Hi$ Fixed-Order Optimization) to various
benchmarks for fixed-order $\Hi$ controller design. The plants in
the examples are all finite-dimensional, linear time-invariant and
multi-input-multi-output (MIMO). The controller order is fixed
\emph{a priori} to be less than the order of plant.   The design
problem is to minimize the $\Hi$ norm of the transfer function for
the closed loop plant.  This is a difficult optimization problem due
to the non\-convexity and non\-smoothness of the objective function.
\hifoo\ uses a hybrid algorithm for nonsmooth, nonconvex optimization
based on several techniques to attempt to find fixed-order
controllers achieving the minimum closed-loop $\Hi$ norm.  The
results are compared with published results using other techniques.

Benchmark examples are chosen from both applied and academic
test problems:
\begin{enumerate}
  \item \textbf{AC$8$:} A $9^{\textrm{th}}$-order state-space model of the linearized vertical plane dynamics of an aircraft \cite{TransportPlane};
  \item \textbf{HE$1$:} A $4^{\textrm{th}}$-order model of the longitudinal motion of a
  VTOL helicopter for typical loading and flight condition at the
  speed of $135$ knots \cite{VTOL}, and \textbf{VTOL}, a variation of this model;
  \item \textbf{REA$2$:} A $4^{\textrm{th}}$-order chemical reactor model
  \cite{Chemicalreactor}, and \textbf{CR}, a variation of this model;
  \item \textbf{AC$10$:} A $55^{\textrm{th}}$-order aeroelastic model
  of a modified Boeing B-$767$ airplane at a
  flutter condition \cite{AC10example};
  \item \textbf{BDT$2$:} An $82^{\textrm{nd}}$-order realistic model of a binary distillation
  tower with pressure variation considered in model description
  \cite{BDTexample};
  \item \textbf{HF$1$:} A $130^{\textrm{th}}$-order one-dimensional model for heat flow
  in a thin rod \cite{HFexample};
  \item \textbf{CM$4$:} A $240^{\textrm{th}}$-order cable
  mass model describing a hybrid parameter system representing
  nonlinear dynamic response of a relief valve used to protect a
  pneumatic system from overpressure \cite{CMexample};
  \item \textbf{PA:} A $5^{\textrm{th}}$-order model of a
  piezoelectric bimorph actuator system \cite{Piezoelectric};
  \item \textbf{HIMAT:} A $20^{\textrm{th}}$-order model of
  an experimental highly maneuverable (HIMAT) airplane which is a scaled and remotely
  piloted version of an advanced fighter \cite{HIMATExGod};
  \item \textbf{VSC:} A $4^{\textrm{th}}$-order quarter-car model consisting of one-fourth of
  the body mass and suspension components of a car and one wheel. This model is used extensively
 in the literature and captures many essential characteristics of a real suspension system;
  \item \textbf{AUV:} This linear model of a cruise control system is obtained by
  linearizing the non-linear model of an autonomous underwater vehicle, Subzero III, around its cruising
  condition. Three SISO autopilots (speed, heading and depth autopilots) need to be developed for the flight control of
  Subzero III. The plant models for speed, heading and depth
  autopilots are $3^{\textrm{rd}}$,
  $5^{\textrm{th}}$ and $6^{\textrm{th}}$-order respectively
  \cite{AUVEx};
  \item \textbf{Enns' Example:} This $8^{\textrm{th}}$-order plant was proposed by
  D.~F.~Enns \cite{Enns_example}. This example is used as an academic test problem in the literature for designing reduced-order $\Hi$ controllers;
 \item \textbf{Wang's Example:} This $4^{\textrm{th}}$-order plant
 was suggested by J.-Z.~Wang as a theoretical benchmark problem in \cite{Wang2003},
Example~$6.2$.
\end{enumerate}

Note that benchmark examples $1-11$ are taken from real applications
and $12-13$ are academic test problems. The problem data for
examples $1-8$ are obtained from the \compleib $\;$library
\cite{Compleib} and those for examples $9-13$ are collected from
various papers in the literature.  For another collection of
results using \hifoo, see \cite{Henrion2006}.

The rest of the paper is organized as follows.
The problem of fixed-order $\Hi$ controller design is
described and the optimization method used by \hifoo\
is summarized in Section \ref{problemandopt}.
Our computational results and comparisons with those published
for other methods are given in Section \ref{examples}.
Concluding remarks are in Section \ref{concluding}.

\section{Problem Formulation and Optimization Method} \label{problemandopt}
\begin{figure}
\begin{center}
\begin{picture}(100,100)(0,30) \label{fig:gensys}
\thicklines \put(30,100){\framebox(40,40)[c]{$G(s)$}}
\put(30,30){\framebox(40,40)[c]{$K(s)$}}
\put(30,130){\vector(-1,0){40}} \put(-20,130){$z$}
\put(110,130){\vector(-1,0){40}} \put(116,130){$w$}
\put(30,110){\line(-1,0){20}} \put(10,110){\line(0,-1){50}}
\put(10,60){\vector(1,0){20}} \put(0,80){$y$}
\put(90,110){\vector(-1,0){20}} \put(90,110){\line(0,-1){50}}
\put(70,60){\line(1,0){20}} \put(95,80){$u$}
\end{picture}\\
\caption{Standard Feedback System}
\end{center}
\end{figure}
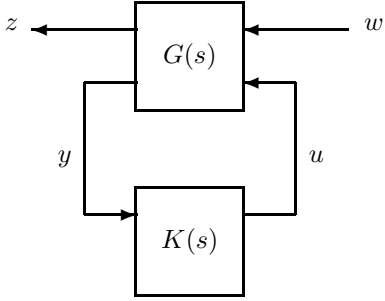
Consider the standard feedback system with generalized plant, $G$,
with state space realization \be \label{eq:plant}
G(s)=\left[\begin{array}{c|cc}
  A & B_{1} & B_2 \\
  \hline
  C_{1} & D_{11} & D_{12} \\
  C_2 & D_{21} & D_{22} \\
\end{array}\right],
\ee where $A\in\mathcal{R}^{n\times n}$,
$D_{12}\in\mathcal{R}^{p_1\times m_2}$,
$D_{21}\in\mathcal{R}^{p_2\times m_1}$ and other matrices have
compatible dimensions. Let the controller have state space
realization $K(s)=\left[\begin{array}{c|c}
  A_K & B_K \\
  \hline
  C_K & D_K \\
\end{array}\right],$ where $A_K\in\mathcal{R}^{n_K\times n_K}$ and
$B_K, C_K, D_K$ have dimensions that are compatible with $A_K$ and
the plant matrices.
The transfer function from the input $w$ to output $z$ is
\bea
\mathcal{F}_l(G,K)=G_{11}+G_{12}(I-G_{22}K)^{-1}G_{21}
\eea
where $G_{ij}(s)=C_i(sI-A)^{-1}B_j+D_{ij}$ for $i, j=1,2$.

The optimal $\Hi$ controller design can be formulated as
minimization of the closed loop $\Hi$ norm function
\bea
\inf_{K~\textrm{stabilizing}} ||\mathcal{F}_l(G,K)||_\infty  ,
\eea
where $K$ internally stabilizes the closed-loop system. When the
controller order $n_K$ equals the plant order $n$, methods are
known to compute the controller that minimizes the $\Hi$ norm.
However, unless $n$ is very small, implementation of full-order
controllers is generally not practical or desirable.

For this reason, we consider the same problem
with the controller order $n_K$ fixed to a number smaller than $n$.
We refer to this as the \textit{Fixed-Order $\Hi$ Controller Design} problem.
The closed-loop $\Hi$ norm function is, in general, nonconvex
and nonsmooth, and often is not differentiable at local minimizers.
The stability constraint is also nonconvex and nonsmooth.
Thus, no method is known for finding a guaranteed global minimum.
\hifoo\ uses a two-stage approach, stabilization followed by performance
optimization.  In the first stage, \hifoo\ proceeds to minimize the
spectral abscissa (maximum of the real parts of the eigenvalues) of
the closed loop system matrix with respect to the free parameters
in the controller, until the spectral abscissa is negative
(a controller has been found that stabilizes the closed loop system).
If no stabilizing controller is found, \hifoo\ terminates with an
error message.  In the second stage, \hifoo\ attempts to locally minimize
the $\Hi$ performance of the closed loop system.  Both stages use \hanso,
a code for nonsmooth, nonconvex optimization with the following elements:
\begin{itemize}
  \item a quasi-Newton algorithm (BFGS) initial phase provides a fast way to approximate
a local minimizer;
  \item a local bundle phase attempts to verify local optimality for
the best point found by BFGS, and if this does not succeed,
  \item a gradient sampling phase \cite{BurkeSIAM05},
  \cite{BurkeTAC06}
attempts to refine the approximation of the local minimizer, returning a rough
optimality measure.
\end{itemize}
The last two phases are invoked only if the quadratic programming
solver {\tt quadprog} is installed; see below.
All three of these optimization techniques use gradients which are
automatically computed by \hifoo. No effort is made to identify the
exceptional points where the gradients fail to exist. The algorithms
are not defeated by the discontinuities in the gradients at
exceptional points. The BFGS phase builds a highly ill-conditioned
Hessian approximation matrix, and the bundle and gradient sampling
final phases search for a point in parameter space for which a
convex combination of gradients at nearby points has small norm.
More details are given in \cite{BHLO-IFAC-06}.  We used \hifoo\ 1.5
\cite{Millstone2006}, which differs from \hifoo\ 1.0
\cite{BHLO-IFAC-06} in that in \hifoo\ 1.5, the $D_{22}$ block is
allowed to be nonzero and specification of a sparsity pattern for
the controller is possible. However, we did not make use of these
features; $D_{22}$ is zero for all the examples below.

\hifoo\ is freely available \matlab\ code$^1$\footnotetext[1]{
http://www.cims.nyu.edu/overton/software/hifoo/}
and has been designed to be easy to use. It is built on the \hanso\
optimization package, freely available at the same web site. It does
not require any external software beyond the \matlab\ Control System
Toolbox, but it runs much faster if the {\tt linorm} function of the
\slicot\ package is installed and in the \matlab\ path (available
commercially from {\tt www.slicot.de}, but freely available from the
\hifoo\ web page for noncommercial use with \hifoo\ using \matlab\
running under Windows).  \hifoo\ also makes use of the {\tt
quadprog} quadratic programming solver from \mosek\ or the \matlab\
Optimization Toolbox if it is installed and in the \matlab\ path,
but this is not required. Our experiments used \matlab\ 2006a with
{\tt linorm} and {\tt quadprog} installed.

Because \hifoo\ uses randomized starting points, and also the gradient
sampling phase involves randomization, the same results are not obtained
every time \hifoo\ is run.  For this reason,
each result reported below is the minimum closed-loop $\Hi$ norm found
in $10$ runs for each fixed controller order for each benchmark example.
For moderate size problems (plant order $1-20$) and low-order controllers
(order $0-4$), the running time typically required for one run of \hifoo\
is on the order of a few seconds.  All the running times were limited
to 5 minutes by setting the option {\tt options.cpumax} to 300 seconds.
More details on the times required are given in a report available
on the web${^2}$.\footnotetext[2]{
http://www.cims.nyu.edu/overton/papers/pdffiles/acc08times.pdf}

\section{Results on Benchmark Problems}
\label{examples}
\subsection{Examples from the \compleib\ Library}
In \cite{AN-TAC-06}, nonsmooth $\Hi$ synthesis algorithms are described
and tested on various synthesis problem from the \compleib\ library
\cite{Compleib}. The philosophy of using direct nonsmooth optimization
is similar to ours but the algorithmic details are different.
Fixed-order $\Hi$ controllers are designed for the
problems and the performance of the nonsmooth $\Hi$ algorithm is
compared with a specialized augmented Lagrangian algorithm
\cite{ANT-JRNC-03}, the Frank-Wolfe algorithm \cite{GOA-TAC-97} and
full-order $\Hi$ controller design method by the DGKF technique
\cite{DGKF}.

In the results given in \cite{AN-TAC-06},
the nonsmooth $\Hi$ algorithm performs best for
all benchmark problems except the plant REA$2$ for which the augmented
Lagrangian algorithm gives a better result. We applied \hifoo\ to the
same benchmark examples and compared our results with the augmented
Lagrangian result for plant REA$2$ and the nonsmooth $\Hi$ results
for the other examples. The results are given in Table~\ref{compleibtable}.
The third and fourth columns display the final value of the
$\Hi$ norm for the closed-loop plant along with the controller order,
comparing the results from \cite{AN-TAC-06} with the results using
\hifoo.  For comparison, the second column shows the $\Hi$ norm for the
closed-loop system using an optimal full-order controller.

\begin{table}[h]
\caption{Comparison on Examples from the \compleib\ Library}
\label{compleibtable} \vspace{-6mm}
\begin{center}
\begin{tabular}{|c|c|c|c|}
  \hline
  \hline
   &  \multicolumn{3}{c|}{}\\
   &  \multicolumn{3}{c|}{$\left(\|\mathcal{F}_l(G,K)\|_\infty, n_K\right)$}\\
  \cline{2-4}
   &  & & \\
  Plant & Full-Order& \cite{AN-TAC-06} & \hifoo\ \\
  \hline
   &  &  &    \\
  AC$8$ & $(1.892, 9)$ & $(2.005, 0)$ & $(2.005, 0)$ \\
  HE$1$ & $(0.0737, 4)$ & $(0.154, 0)$ & $(0.154, 0)$ \\
  REA$2$ & $(1.135, 4)$ & $(1.155^{\dag}, 0)$ & $(1.149, 0)$ \\
  AC$10$ & $(3.23, 55)$ & $(13.11, 0)$ & $(12.83^*, 0)$ \\
  AC$10$ & $(3.23, 55)$ & $(10.21, 1)$ & $(10.338^*, 1)$ \\
  BDT$2$ & $(0.234, 82)$ & $(0.8364, 0)$ & $(0.6515, 0)$ \\
  HF$1$ & $(0.447,130)$ & $(0.447, 0)$ & $(0.447, 0)$\\
  CM$4$ & $(0.816, 240)$ & $(0.816, 0)$ & $(0.816, 0)$ \\
  \hline
  \hline
\end{tabular}
\end{center}
$^{\dag}$ Augmented Lagrangian method \hfill $^*$ Stable Starting
Point
\end{table}

As seen in Table \ref{compleibtable}, \hifoo\ gives better performance
than other algorithms for plants REA$2$ and BDT$2$ and the same
performance for plants AC$8$, HE$1$, HF$1$ and CM$4$.
Using its default randomly generated starting conditions,
\hifoo\ has difficulty finding a stabilizing controller for AC$10$, because
of the very different scalings of the variables.
Therefore, we provided an initial stable starting point from \cite{BurkeSIAM05}.

Note that both \cite{AN-TAC-06} and \hifoo\ find that, for
the high-order plants HF$1$ and CM$4$, full-order controller performance
can actually be achieved by static output feedback.  This interesting
observation shows the value of the optimization approach.

\subsection{Comparison with $\Hi$ Multidirectional Search Method}
${}$\\ We consider static output-feedback $\Hi$ synthesis for the
plants VTOL Helicopter (VTOL), Chemical Reactor (CR) and
Piezoelectric Actuator (PA). The first two are slight variations on
HE$1$ and REA$2$, respectively. The state-space data for these examples
are taken from \cite{ApkarianTVCP} to use the same data set as
\cite{ApkarianMultiDirectional}.

An algorithm combining multidirectional search (MDS) with nonsmooth
optimization techniques is given in \cite{ApkarianMultiDirectional}.
The algorithm is applied to the plants above for static
output-feedback $\Hi$ synthesis and its results compared with
the Augmented Lagrangian method (AL) described in \cite{ApkarianTVCP}.
We applied \hifoo\ to the same problems and the results are given in
Table \ref{staticHinf}.

\begin{table}[h]
\caption{Comparison with Multidirectional Search Method}
\label{staticHinf} \vspace{-6mm}
\begin{center}
\begin{tabular}{|c|c|c|c|}
  \hline
  \hline
   &  \multicolumn{3}{c|}{}\\
   &  \multicolumn{3}{c|}{$\left(\|\mathcal{F}_l(G,K)\|_\infty, n_K\right)$}\\
  \cline{2-4}
   &  & & \\
  Plant & Full-Order & \cite{ApkarianMultiDirectional} & \hifoo\\
  \hline
   &  &  & \\
  VTOL & $(0.0737, 4)$ & $(0.157, 0)^{\dag}$ & $(0.154, 0)$ \\
  CR & $(1.135, 4)$ & $(1.183, 0)$ & $(1.168, 0)$ \\
  PA & numerically & $(1.76e{-4}, 0)$& $(1.18e{-4}, 0)$ \\
 & ill-posed & & \\
  \hline
  \hline
\end{tabular}
\end{center}
$^{\dag}$ Augmented Lagrangian method
\end{table}

The controllers obtained by \hifoo\ for static-output feedback $\Hi$
synthesis have lower closed-loop $\Hi$ cost compared to other
methods for the benchmark problems above.

\subsection{Enns' Benchmark Problem}
We consider fixed-order $\Hi$ controller design of a plant proposed by
Enns \cite{Enns_example}. This example is used as a benchmark
problem in the literature to design reduced-order $\Hi$ controllers.
The optimal $\Hi$ norm achieved in closed-loop by a full-order (order 8)
controller is $1.1272$.

In \cite{Enns_Survey_Zhou}, several controller reduction methods are
compared, including weighted additive and coprime factor controller
reduction methods, and these are applied to Enns' benchmark problem.
In \cite{Enns_ex2001} and \cite{Enns_ex2002} reduced-order controllers
are obtained by weighted $\Hi$ model reduction and a block-balanced
truncating algorithm respectively.  Recent enhancements of several
frequency-weighted balancing related controller reduction methods
are discussed in \cite{Enns_ex2005}.

We applied \hifoo\ to the same benchmark example and compare the
results with those obtained in \cite{Enns_Survey_Zhou}
as well as by the other methods \cite{Enns_ex2001},
\cite{Enns_ex2002}, \cite{Enns_ex2005} in Table \ref{ennstable}.
For all of orders 1 through 7, \hifoo\ finds controllers with lower
closed-loop  $\Hi$ norm. Therefore, the performance of \hifoo\ is
better than other methods for this particular benchmark problem.
Note that while the other methods compute a full-order controller
first and then apply techniques to reduce its order, \hifoo\ does
not compute a full-order controller, but computes low-order
controllers directly.

\begin{table}[h]
\caption{Comparison on Enns' Example} \label{ennstable}
\vspace{-4mm}
\begin{center}
\begin{tabular}{|c|c|c|c|c|c|}
  \hline
  \hline
  & \multicolumn{5}{c|}{}\\
   & \multicolumn{5}{c|}{$\|\mathcal{F}_l(G,K)\|_\infty$}\\
   \cline{2-6}
   & &  & & &\\
 $n_K$ & \cite{Enns_Survey_Zhou} & \cite{Enns_ex2001} & \cite{Enns_ex2002} & \cite{Enns_ex2005} & \hifoo\  \\
  \hline
   &  &  &  & & \\
  $7$ & $1.1960$ & $1.1957$ & $1.198$ & 1.1950 & $1.1655$ \\
  $6$ & $1.1960$ & $1.1971$ & $1.196$ & 1.1960 & $1.1447$ \\
  $5$ & $1.1950$ & $1.1970$ & $1.204$ & 1.1960 & $1.1508$ \\
  $4$ & $1.1950$ & $1.1991$ & $1.197$ & 1.1960 & $1.1923$ \\
  $3$ & $1.4880$ & $1.8801$ & $3.906$ & 2.7580 & $1.1921$ \\
  $2$ & $1.4150$ & $1.9681$ & $1.954$ & 1.4130 & $1.2438$ \\
  $1$ & $2.4670$ & $73.2860$ & Unstable & Unstable & $1.4256$\\
  \hline
  \hline
\end{tabular}
\end{center}

\end{table}

\subsection{HIMAT Example}
Longitudinal dynamics of an experimental highly maneuverable (HIMAT)
airplane make a well-known benchmark example for reduced-order robust
controller design \cite{HIMATExGod, SreeramHIMAT}. The generalized
plant has $20$ states and the optimal $\Hi$ norm achieved in
closed-loop by a full-order controller is $0.9708$.

The controller reduction techniques in \cite{HIMATExGod,
SreeramHIMAT} use frequency-weighted model reduction preserving
$\Hi$ performance. We applied \hifoo\ to the HIMAT example as an
alternative to controller reduction. The results can be seen in
Table \ref{HIMATTable}.

\begin{table}[h]
\caption{Comparison on HIMAT Example} \label{HIMATTable}
\vspace{-4mm}
\begin{center}
\begin{tabular}{|c|c|c|c|}
  \hline
  \hline
  & \multicolumn{3}{c|}{}\\
   & \multicolumn{3}{c|}{$\|\mathcal{F}_l(G,K)\|_\infty$}\\
   \cline{2-4}
   & &  & \\
 $n_K$ & \cite{HIMATExGod} & \cite{SreeramHIMAT} & \hifoo\  \\
  \hline
   &  &  &  \\
  $16$ & $0.98$ & $0.97$ & $1.01$ \\
  $15$ & $-$ & $0.97$ & $1.01$ \\
  $14$ & $-$ & $0.97$ & $1.01$ \\
  $13$ & $0.98$ & $0.98$ & $1.01$ \\
  $12$ & $-$ & $0.98$ & $1.01$ \\
  $11$ & $-$ & $0.99$ & $1.02$ \\
  $10$ & $2.02$ & $1.27$ & $1.03$\\
  $7$ & $1.27$ & $1.22$ & $1.06$\\
  $6$ & $-$ & $1.22$ & $1.07$\\
  \hline
  \hline
\end{tabular}
\end{center}

\end{table}

Note that \hifoo\ gives better performance compared to other methods when the controller order is low.
When the controller order is close to the plant order, other methods perform better.
However, the difference between performance is small.
This example shows that although \hifoo\ gives good results when controller order is high,
its best results are obtained when the controller order is small
which is the case in almost all practical implementations.

\subsection{Vehicle Suspension Control (VSC)}
A simple quarter-car suspension model consists of one-fourth of the
body mass and suspension components and one wheel. The model has $4$
states and captures essential characteristics of a real suspension
system. The suspension system is controlled by a hydraulic
actuator for ride comfort, road holding ability and suspension
deflection.  An $\Hi$ control problem is formulated by weighting
three different objectives for vehicle suspension \cite{VSC1997}.

In \cite{VSC2005}, a static output feedback $\Hi$ controller for the
quarter-car suspension model with semi-active damper is obtained
using a genetic algorithm. Table~\ref{VSCTable} shows the comparison
between \cite{VSC2005} and \hifoo. Note that \hifoo\ finds a static
$\Hi$ controller achieving closed-loop $\Hi$ norm close to the
optimal value for a fourth-order controller.

\begin{table}[h]
\caption{Comparison on Vehicle Suspension Control Example}
\label{VSCTable} \vspace{-4mm}
\begin{center}
\begin{tabular}{|c|c|c|c|}
  \hline
  \hline
    & \multicolumn{3}{c|}{}\\
    & \multicolumn{3}{c|}{$\|\mathcal{F}_l(G,K)\|_\infty$}\\
    \cline{2-4}
   & &  & \\
 Plant & Full-Order & \cite{VSC2005} & \hifoo\  \\
  \hline
   &  &  & \\
   quarter-car suspension model & $3.216$ & $7.640$  & $3.975$\\
    with semi-active damper &  &   & \\
  \hline
  \hline
\end{tabular}
\end{center}
\end{table}

\subsection{Autonomous Underwater Vehicle (AUV)}
In \cite{AUVEx}, autopilots (forward speed, heading and depth) are
designed to control an autonomous underwater vehicle with
performance objectives. It is desirable to have a low-order autopilot
for implementation purposes. Therefore, a reduced-order $\Hi$ control
problem is posed as a rank minimization problem and a
solution is approximated by a trace minimization approach.

Table \ref{AUVTable} shows that \hifoo\ achieves lower closed-loop
$\Hi$ norm with a smaller controller order compared to \cite{AUVEx}.
\begin{table}[h]
\caption{Comparison on Autonomous Underwater Vehicle Example}
\label{AUVTable} \vspace{-4mm}
\begin{center}
\begin{tabular}{|c|c|c|c|}
  \hline
  \hline
    & \multicolumn{3}{c|}{}\\
    & \multicolumn{3}{c|}{$\left(\|\mathcal{F}_l(G,K)\|_\infty, n_K\right)$}\\
    \cline{2-4}
   & & &  \\
Autopilots & Full-Order & \cite{AUVEx} & \hifoo\ \\
  \hline
   &  & &   \\
  Speed & $(0.9538,3)$  & $(0.9550,1)$ & $(0.9543,1)$ \\

   &  & &   \\
  Heading & $(0.9536,5)$ & $(0.9633,3)$ & $(0.9540,2)$ \\
   &  & & $(0.9545,1)$  \\
   &  & & $(0.9548,0)$  \\

   &  &  &  \\
  Depth & $(0.9556,6)$ & $(0.9798,3)$  & $(0.9621,1)$\\
  \hline
  \hline
\end{tabular}
\end{center}
\end{table}

\subsection{Wang's Example}
We consider the theoretical example in \cite{Wang2003},
Example~$6.2$. Controller approximation approaches preserving $\Hi$ performance are suggested in \cite{HIMATExGod}.
The $\Hi$ controller reduction problem is converted to a frequency weighted model reduction problem. The
controller reduction method in \cite{HIMATExGod} is generalized in \cite{Wang2003}.

In \cite{Wang2006}, algorithms based on a cone complementarity
linearization idea are proposed to solve the nonconvex feasibility
problems for controller order reduction. The results are compared
with \cite{Wang2003} and better performance is observed. We applied
\hifoo\ to the same problem and the results are shown in Table
\ref{WangTable}. The closed-loop $\Hi$ norms for \cite{Wang2003} and
\cite{Wang2006} are computed using the controllers shown in
the corresponding papers and are less than the theoretical upper bounds
in the papers. Note that the controllers found by \hifoo\ give
closed-loop $\Hi$ norm close to the result for a full-order
controller.

\begin{table}[h]
\caption{Comparison on Wang's Example} \label{WangTable}
\vspace{-4mm}
\begin{center}
\begin{tabular}{|c|c|c|c|}
  \hline
  \hline
   \multicolumn{4}{|c|}{}\\
   \multicolumn{4}{|c|}{$\left(\|\mathcal{F}_l(G,K)\|_\infty, n_K\right)$}\\
    \cline{1-4}
   & & &  \\
 Full-Order& \cite{Wang2003} & \cite{Wang2006} &  \hifoo\  \\
  \hline
   &  & &   \\
  $(50.640,4)$ & $(55.621,3)$  & $(58.096,3)$ & $(50.642,2)$ \\
   & $(55.639,2)$ & $(55.624,2)$ &  $(50.645,1)$ \\
   &  & & $(50.879,0)$ \\
  \hline
  \hline
\end{tabular}
\end{center}
\end{table}

\section{Concluding Remarks} \label{concluding}
In this note, we reported on results of applying the \hifoo\ Toolbox to
various benchmark problems for fixed-order and reduced-order $\Hi$
design. The examples were mostly chosen from various applications and
also included two academic test problems.

The performance of \hifoo\ is better compared to existing results in
the literature in most cases. We conclude that \hifoo\ is an
effective alternative method for fixed-order $\Hi$ controller
design.
\section{Acknowledgments}
The authors would like to thank P.~J.~Goddard for supplying the data for the HIMAT example,
V.~Sreeram for providing the HIMAT example code and P.~Apkarian for valuable comments.
We particularly thank M.~Millstone for developing version 1.5 of \hifoo.
The work of the second author was supported in part by the
U.S. National Science Foundation Grant DMS-0714321. The views expressed in this
paper are those of the authors and are not necessarily shared by the
NSF.

\end{document}